\documentclass[12pt]{article}

\usepackage{amsmath,amssymb, amsthm}

\begin{document}

\title{Design of an Experiment to Test Nonclassical Probabilistic Behavior of the Financial market}
\author{Andrei Khrennikov\\
International Center for
Mathematical Modeling \\ in Physics and Cognitive Sciences\\
University of V\"axj\"o, S-35195, Sweden}

\maketitle

\abstract{The recent crash demonstrated (once again) that the
description of the financial market by  present financial
mathematics cannot be considered as totally satisfactory. We
remind that nowadays financial mathematics is heavily  based  on
the use of random variables and stochastic processes which are
described by Kolmogorov's measure-theoretic model for probability
(``classical probabilistic model''). I speculate that the present
financial crises is a sign (a kind of experiment to test validity
of classical probability theory at the financial market) that the
use of this model in finances should be either totally rejected or
at least completed. One of the best candidates for a new
probabilistic financial model is quantum probability or its
generalizations, so to say quantum-like (QL) models. Speculations
that the financial market may be nonclassical have been present in
scientific literature for many years. The aim of this note is to
move from the domain of speculation to rigorous statistical
arguments in favor of probabilistic nonclassicality of the
financial market. I design a corresponding statistical test which
is based on violation of the formula of total probability (FTP).
The latter is the basic in classical probability and its violation
would be a strong sign in favor of QL behavior at the market.}

\section{Introduction}

Detailed review on applications of quantum mathematics to finances
can be found in author's monograph \cite{KHR}, we mention just
some publications \cite{KHR}--\cite{BBL}.

The aim of this paper was formulated in the abstract. We  point
out that the experimental test to check a possibility of violation
of FTP at the financial market can be considered as adaptation to
finances of the general statistical test proposed in \cite{test}.
Its version was already tested in cognitive science, see Conte et
al. \cite{COU}, \cite{COU1}. It was shown that FTP (and hence
classical probability theory) is violated in some experiments on
recognition of ambiguous pictures.

Our experiment may be criticized by dealers working at the real
market. We cannot exclude such a possibility. However, our
experiment opens the door toward designing of similar may be more
realistic financial experiments. As a first step, one may try to
perform our experiment with students.

\section{Supplementary (``Complementary'')  Stocks}

In ordinary QM one pays a lot of attention to so called {\it
complementary observables}; for example, position and momentum.
Since we will operate with discrete observables, we can mention
electron's spin projections to two different directions or
photon's polarization projections as examples of complementary
observables. Complementary observables are represented in QM
formalism by noncommuting operators.

In QL-finances, we will also operate with observables which are
represented by noncommuting operators.

The delicate point is that, unlike QM, at the moment we do not
have a quantization procedure for financial variables -- to
produce from variables operators. In a series of papers, e.g.,
\cite{KJ}, books \cite{KHR}, \cite{BOOK}, I developed a kind of
quantization procedure: starting with probabilities one can
produce operator representation of variables. Therefore it would
be useful to formulate in pure probabilistic terms condition of
noncommutativity.

Consider two observables $a$ and $b$ in QM. Suppose that they are
dichotomous, $a \in X_a=\{ \alpha_1, \alpha_2\}$ and  $b \in
X_b=\{\beta_1, \beta_2\}.$ They can be represented in QM formalism
by self-adjoint operators in the two dimensional complex space,
${\cal H}_2={\bf C} \times {\bf C};$ thus by $2\times 2$ Hermitian
matrices: $\widehat{a}, \widehat{b}.$ We recall that any symmetric
matrix can be diagonalized in the basis consisting of its
eigenvectors, say
$$
\widehat{a} e_\alpha^a= \alpha e_\alpha^a, \; \alpha=\alpha_1,
\alpha_2;
$$
$$
\widehat{b} e_\beta^b= \beta e_\beta^b, \; \beta = \beta_1,
\beta_2.
$$
The crucial point is that if operators do not commute, then they
cannot be diagonalized in the same basis. It means that
$$
\langle e_\alpha^a, e_\beta^b \rangle \not=0,
$$
for any pair of values $\alpha, \beta.$ Conditional (transition)
probabilities can be expressed via scalar products of
corresponding eigenvectors:
$$
{\bf P}(b=\beta \vert a=\alpha)= \vert \langle e_\alpha^a,
e_\beta^b \rangle \vert^2.
$$
Thus two observables are complementary iff all these probabilities
are strictly positive:
\begin{equation} \label{SC} {\bf P}(b=\beta \vert a=\alpha) > 0
\end{equation}
for all $\alpha$ and $\beta.$ The latter condition has no direct
relation to QM. It can be used in any domain of science.

We now analyze little bit the meaning of this condition. It can be
equivalently written as
\begin{equation}
\label{SCa}{\bf P}(b=\beta \vert a=\alpha) \not= 1
\end{equation}
Thus it is impossible to determine a value $b=\beta$ by fixing the
value $a=\alpha.$ The $b$-variable has some features which cannot
be determined on the basis of features of the $a$-variable. Thus
$b$ contains additional, or to say supplementary,
information.Therefore we call observables (from any domain of
science) {\it supplementary} if (\ref{SC}) holds. We may call them
complementary as Bohr did in QM. But, unlike Bohr, we do not
emphasize {\it mutual exclusivity of measurements.} In principle,
supplementary observables, unlike complementary, can be measured
simultaneously. However, supplementary observables are also
mathematically  represented by noncommuting operators. We recall
that our aim is to show the adequacy of the mathematical apparatus
of QM to the financial market. Thus, we need not borrow even
quantum ideology and philosophy.

We will consider supplementary stocks. Formally, one can determine
either two stocks, say $A$ and $B,$ are supplementary or not by
using the formal definition (\ref{SC}).  However,  to do this, we
should perform experiment for a large ensemble of dealers. If,
finally, one observes that transition frequencies are close to
zero, it will imply that this pair of stocks is not useful for
coming interference experiment. Therefore it is much better to use
financial intuition to determine either two stocks are intended to
be supplementary or not.

\section{Experiment Design}

1). Select two supplementary stocks, say $A$ and $B.$

2). Select of an ensemble $\Omega$ of dealers who used to work
with these two stocks. Its size $N$ should be large enough

3). Select an interval $\delta$ giving average time between two
 successive financial operations.\footnote{If during some period
 of time $T$ (e.g., depending of frequency of operating, one day,
 or month, or year), a dealer  made $k$ operations at the financial
 market, then  $\delta$ is equal to average of $T/k$ with respect
 to all dealers from the ensemble $\Omega$ selected for the
 experiment.}

4). Define two observables: for a dealer $\omega \in \Omega,$
$a(\omega)=+1$ if he has bought a packet of stocks during the
period $\delta$ and $a(\omega)=-1$ if he has not.\footnote{Even if
his $A$-bid was present at the market, but it did not match asked
prices for this stock; $a=-1$ as well if he did not submit any
$A$-bid.} The $b$-observable is defined in the same way.

4). Starting with some initial instant of time, say $t_0,$ wait
until $t_0+\delta.$ Then count all dealers who has bought during
this period some packet\footnote{In the experiment under
consideration the size of packet does not play any role. However,
experiment can be design in a more complicated way, by including
the size of a packet. In this way nonsignificant bids can be
excluded from the game.} of $A$-stocks, i.e., all elements $\omega
\in \Omega$ such that $a=+1.$ Denote this number by $n^a(+).$ We
define frequency probabilities
$$
p^a(+)= n^a (+)/N, p^a(-)=1 - p^a(+);
$$
In the same way we find $n^b(+)$ -- the number of dealers whose
$B$-bids has matched asks at the market (during the same period
$[t_0, t_0+\delta]$ -- and define frequency probabilities
$p^b(\pm).$

5). On the basis of previous $a$-measurement select from $\Omega$
sub-ensembles of dealers $\Omega^a_+$ -- those whose $A$-bids were
realized during the period $[t_0, t_0+\delta]$ -- and $\Omega^a_-$
-- those whose $A$-bids did not match any asked price for the
$A$-stock or those who did not bid anything for this stock. Denote
the numbers of elements in these ensembles by $N^a_+$ and $N^a_-,$
respectively.

6). Wait until $t_0+2\delta$ and after this count all dealers from
$\Omega^a_+$ whose $B$-bids were realized  during the period
$[t_0+\delta, t_0+2\delta].$  These are elements $\omega\in
\Omega^a_+$ for whom $b(\omega)=+1.$ Denote obtained number by
$n(+ \vert +).$  We define frequency probabilities
$$
p(+ \vert +)= n(+ \vert +)/N^a_+,  \; \;  p(- \vert +)=1 -
p(+\vert +).
$$
They have the meaning of conditional probabilities. For example,
$p(+ \vert +)$ is probability that a randomly chosen dealer first
bought a packet of the $A$-stocks and then a packet of the
$B$-stocks.

 In  same way we define frequency probabilities
$$p(+\vert -)= n(+ \vert -)/N^a_-, \; \; p(-\vert -)=1 - p(+ \vert -)$$
by making the $b$-measurement for dealers belonging to the
sub-ensemble $\Omega^a_-.$

6). Finally, define the  coefficient
\begin{equation}
\label{FYT} \lambda_\beta= \frac{p^b(\beta)- [p^a(+) p(\beta \vert
+) - p^a(-) p(\beta \vert  -)]}{2\sqrt{p^a(+) p(\beta \vert
+)p^a(-) p(\beta \vert  -)}}, \; \mbox{where}\; \beta= \pm.
\end{equation}

It gives a measure of deviation from the classical formula of
total probability, see Section \ref{FTP}. In quantum mechanics
this coefficient has the form
$$
\lambda_\beta= \cos \theta_\beta
$$
where $\theta_\beta$ is the phase angle. Therefore it can be
called {\it interference coefficient.}

7). An empirical situation with $\lambda \not =0$ would yield
evidence for QL behavior of the financial market: interaction of
dealers and stocks. In this case, starting with the
(experimentally calculated) coefficient of interference $\lambda$
we can proceed either to the conventional Hilbert space formalism
(if this coefficient is bounded by 1) or to so called hyperbolic
Hilbert space formalism (if this coefficient is larger than  1),
see \cite{test}, \cite{KJ} and more in coming book \cite{BOOK}.

\section{Formula  of Total Probability with Interference Term}
\label{FTP}

 In the above notations the conventional
(``classical'') formula of total probability (FTP) is written as
\begin{equation}
\label{F} p^b(\beta)= p^a(+) p(\beta \vert +) - p^a(-) p(\beta
\vert -).
\end{equation}
Thus the probability $p^b(\beta)$ can be found on the basis of
conditional probabilities $p(\beta \vert \pm).$\footnote{ ``The
prior probability to obtain the result e.g. $b=+$ is equal to the
prior expected value of the posterior probability of $b=+$ under
conditions $a=\pm.''$} FTP plays the fundamental role in modern
science. Its consequences are strongly incorporated in modern
scientific reasoning. It is derived in classical probability
theory by using Kolmogorov measure-theoretic model for
probability. This model is the basis of modern financial
mathematics which is based on the use of classical random
variables (and stochastic processes).

In \cite{KHR} I pointed out that the quantum formalism induces a
violation of FTP. An additional term appears in the right hand
side of (\ref{F}), so called {\it interference term.} Violation of
the law of total probability  can be considered as an evidence
that the classical probabilistic description could not be applied.
The $\lambda$-coefficient (\ref{FYT}) gives us a measure of
statistical deviation from FTP.

Our aim is to show that QL probabilistic descriptions could be
applied. The terminology ``quantum-like'' and not simply
``quantum'' is used to emphasize that violations of (\ref{F}) are
not reduced to those which can be described by the conventional
quantum model. In particular, statistical data from the financial
market  may be described by a generalized quantum formalism, see
for details \cite{KHR}, \cite{KJ} and especially \cite{BOOK}.

I would like to than E. Conte and E. Haven for numerous
discussions on QL interference in cognitive and social science. I
also thank M. N. Alonso whose recent Emails, also \cite{MNA},  to
me stimulated me to come back to the problem of QL behavior of the
financial market.

\end{document}